\documentclass{article}
\usepackage{spconf,amsmath,graphicx,hyperref}
\usepackage[acronym]{glossaries}
\usepackage{url}
\usepackage{fancyhdr}
\usepackage{booktabs}
\usepackage{array}
\usepackage{siunitx}
\usepackage{comment}

\title{XSQ-AST: An Explainable Audio Spectrogram Transformer Framework for Localising Synthetic Speech Artifacts}
\name{\vtop{\hbox{
Ben Heritage$^{1,*}$\thanks{$^*$Authors contributed equally.\\
This work was supported by EPSRC Impact Accelerator award EP/X525856/1 and by CoSTARLive Lab, funded by AHRC grant reference AH/Y001079/1.}, 
      Luca Resti$^{1,2,*}$, 
      Mónica Villanueva Aylagas$^{3}$,}\hbox{Timothy Mehlenbacher$^{4}$,
      Konrad Tollmar$^{3}$,
      James Alfred Walker$^{2}$}}}

\address{$^{1}$ AudioLab, School of Physics, Engineering and Technology, University of York, United Kingdom\\
         $^{2}$ Department of Computer Science, University of York, United Kingdom\\
         $^{3}$ SEED -- Electronic Arts (EA), Sweden\\
         $^{4}$ Electronic Arts (EA), United States}

\makeglossaries
\newacronym{ASR}{ASR}{Automatic Speech Recognition}
\newacronym{KDE}{KDE}{Kernel Density Estimation}
\newacronym{MOS}{MOS}{Mean Opinion Score}
\newacronym{SQ-AST}{SQ-AST}{}
\newacronym{AST}{AST}{Audio Spectrogram Transformer}
\newacronym{GradCAM}{GradCAM}{Gradient Class Attribution Maps}
\newacronym{CNN}{CNN}{Convolutional Neural Network}
\newacronym{PPG}{PPG}{Phoneme Posteriorgram}
\newacronym{PDSM}{PDSM}{Phoneme Discretized Saliency Map}
\newacronym{TTS}{TTS}{Text-to-Speech}
\newacronym{XSQ-AST}{XSQ-AST}{eXplainable SQ-AST}

\begin{document}
%\ninept
%
\maketitle

\begin{abstract}
Localising artifacts in synthetic speech remains challenging, as most evaluation methods yield only global quality scores. This paper presents XSQ-AST, a framework that combines the SQ-AST speech quality model with WhisperX phoneme alignment and multiple saliency methods to produce temporally localised artifact diagnostics without model retraining. Saliency maps are projected onto continuous distributions via kernel density estimation and onto phoneme boundaries via phoneme-discretised saliency maps. A 40-participant listening test validated the framework across five perceptual dimensions. Attention Rollout, Attention Flow and an adapted GradCAM produced temporal distributions that correlated with listener highlights, with different methods best suited to different artifact types. An AUC-ROC analysis confirmed discrimination above chance.
\end{abstract}
\begin{keywords}
Speech synthesis, speech processing, saliency detection, feature extraction, quality of experience
\end{keywords}

\section{Introduction}
\label{sec:intro}

The advancement of deep generative architectures has enabled \gls{TTS} systems to synthesise increasingly natural speech. Traditionally, the evaluation of synthetic speech has relied on subjective \glspl{MOS} or objective estimators of such global scores, e.g.~\cite{saeki2022utmos}. 
%NISQA~\cite{mittag2021nisqa} and %

While aggregate metrics provide a useful baseline for assessing overall system performance, they offer little to no interpretability. Furthermore, as recent zero-shot neural codec language models and diffusion-based architectures approach parity with human speech~\cite{chen2024vall, ju2024naturalspeech}, global naturalness scores are becoming increasingly saturated, while highly localised signal degradations remain an open challenge. These local artifacts are comprised of transient temporal discontinuities, localised distortions or momentary prosodic drift due to cumulative error propagation in sequence generation~\cite{neekhara2024improving, yao2025fine}. Localised distortions have also been more broadly investigated in the context of synthetic speech in videogames~\cite{resti2025acoustic}.

To overcome the limited interpretability of global \glspl{MOS}, recent research has pivoted towards more fine-grained approaches. Kuhlmann et al. \cite{kuhlmann2026icassp} introduced an approach to extract frame-level scores that correlate with human perception of localised artifacts. While effective, this solution relies on modifying the training objective with segment-based consistency constraints to compensate for the lack of frame-level quality annotations in standard datasets.

\begin{figure}[t!]
    \centering
    \includegraphics[width=0.9\linewidth]{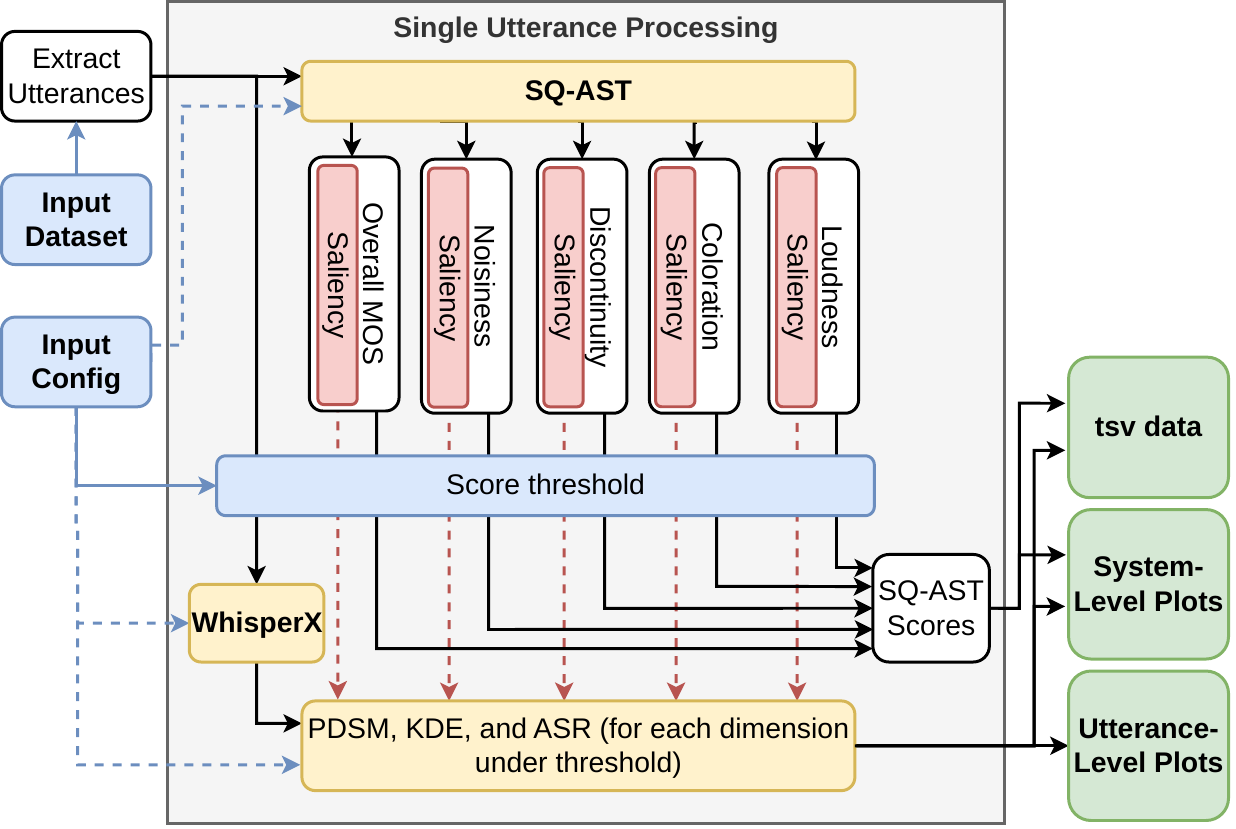}
    \caption{Signal flow for XSQ-AST, containing SQ-AST~\cite{Wardah2025} and WhisperX~\cite{Bain2023} models.}
    \label{fig:methodflow}
\end{figure}

In this paper, \gls{XSQ-AST} is presented as a framework for localising synthesis artifacts directly from pre-trained architectures, avoiding the need for specialised training objectives. By integrating \gls{SQ-AST}~\cite{Wardah2025} with WhisperX~\cite{Bain2023}, the proposed approach extracts feature importance via Attention Rollout~\cite{Abnar2020} and an adapted GradCAM~\cite{Selvaraju2019}. Because \gls{SQ-AST} evaluates perceptual dimensions like noisiness and colouration alongside \glspl{MOS}, degradations are directly associated with specific artifact classes. Time-frequency saliency maps are then projected onto phoneme boundaries to generate \glspl{PDSM}~\cite{Gupta2024}. Finally, this localised information is combined with \gls{KDE}, translating model outputs into explainable diagnostics without requiring model retraining.

\section{Methodology}

The proposed \gls{XSQ-AST} framework acts as a wrapper for SOTA models in speech sound quality estimation~\cite{Wardah2025} and \gls{ASR}~\cite{Bain2023} to gather temporal and frequency-dependent metrics that are interpretable\footnote{Code is available at \url{https://github.com/luca-resti/synth-speech-eval}.}. This section will outline the model input, signal flow as shown in Fig. \ref{fig:methodflow}, and output reports.

\subsection{Input and Preprocessing}

The method takes a dataset of multiple-speaker synthetic speech (in one language) of float or PCM audio files up to the sample rate of \(48\)~kHz, and a configuration file. The configuration file allows the user to specify the settings regarding the score threshold, saliency and \glspl{PDSM}~\cite{Gupta2024}, model settings, \gls{KDE}, and output report types.

The score threshold denotes the threshold for utterances where a full analysis is to be conducted. For large datasets, this both controls the number of reports generated and reduces analysis time. The scores for each of the dimensions (\gls{MOS}, Noisiness, Discontinuity, Colouration, and Loudness) range from \(1-5\), where \(5\) denotes a perfect sample with no artifacts~\cite{Wardah2025}.

% \begin{figure}[t!]
%     \centering
%     \includegraphics[width=1.0\linewidth]{figures/Sample_cutting.drawio.png}
%     \caption{Example utterance splitting procedure}
%     \label{fig:sample_cutting}
% \end{figure}

Preprocessing is conducted to the input dataset by asserting the audio file is above \(2\) seconds, handling multiple channel audio files as separate audio files, resampling audio to target of \(48\)~kHz, and segmenting audio files longer than \(10\) seconds due to constraints in \gls{SQ-AST}. The segmentation works by finding the minimum number of segments with overlap of \(1.5\) seconds for the length of audio, allowing the boundaries of the segmented utterances to be accounted for in both neighbouring utterances. Following steps outlined in ~\cite{Wardah2025}, the utterances are processed by a pre-trained Mel-filter bank feature extractor, and are then normalised to a fixed mean and standard deviation outlined in the \gls{SQ-AST} code.
%(see Fig \ref{fig:sample_cutting} for an example)

\subsection{Saliency Extraction}

When the utterances are parsed through to \gls{SQ-AST}, based upon the \gls{AST} model~\cite{Gong2021}, the method checks if any of the dimension scores are below the threshold outlined in the configuration file, and extracts the saliency via the method requested. The saliency extraction methods of Raw Attention, Attention Rollout, Attention Flow~\cite{Abnar2020}, and an altered \gls{GradCAM}~\cite{Selvaraju2019} are implemented. Raw Attention, Attention Rollout, and Attention Flow are implemented as described in ~\cite{Abnar2020}, whereas \gls{GradCAM} is an adapted version of the method outlined in ~\cite{Selvaraju2019}. While this method is defined for classification tasks in convolutional neural networks, our approach averages the gradients across the last transformer layer, and since it is a regression model, the scoring of \(1\) acts as a proxy for the confidence in class prediction. All of these methods return a saliency map regarding to time-frequency patches in utterances, indicating patches that degrade the score the most. This relies on the assumption that the \gls{SQ-AST} model works by assuming a perfect score, and negating from this with the presence of audio artifacts for a given dimension in the linear layers. The resulting saliency maps are scaled and trimmed to the size of the input Mel-filter bank features, and interpolated.

\subsection{Automatic Speech Recognition}

The utterances that have at least one dimension score under the threshold are downsampled to \(16\)~kHz and concatenated into a large array with padding of one second of silence between utterances to increase efficiency in inference. The WhisperX model~\cite{Bain2023} parses the data to determine the language, transcribe the utterances alongside confidence scores and then align the phonemes from the transcription to the utterances. From the aligned phonemes, we are able to extract the \glspl{PPG}~\cite{hazen2009query} used in the \gls{PDSM} method~\cite{Gupta2024}.

\subsection{Utterance-Level Reports}

\glspl{PPG} and the extracted saliency maps for each dimension are combined by pooling for each phoneme as shown in~\cite{Gupta2024}, to extract the \gls{PDSM}. As well as the pooling methods of sum and mean as shown in~\cite{Gupta2024}, the additional methods of median, max, $\ell_1$ norm, and $\ell_2$ norm are made available to the user. The \gls{PDSM} values are ranked, and the top \(10\%\) of troublesome phonemes by default, in accordance with ~\cite{Gupta2024}, are highlighted in a report alongside the phoneme transcription and Mel-filter feature representation of the speech.

\begin{figure}[t!]
    \centering
    \includegraphics[width=1.0\linewidth]{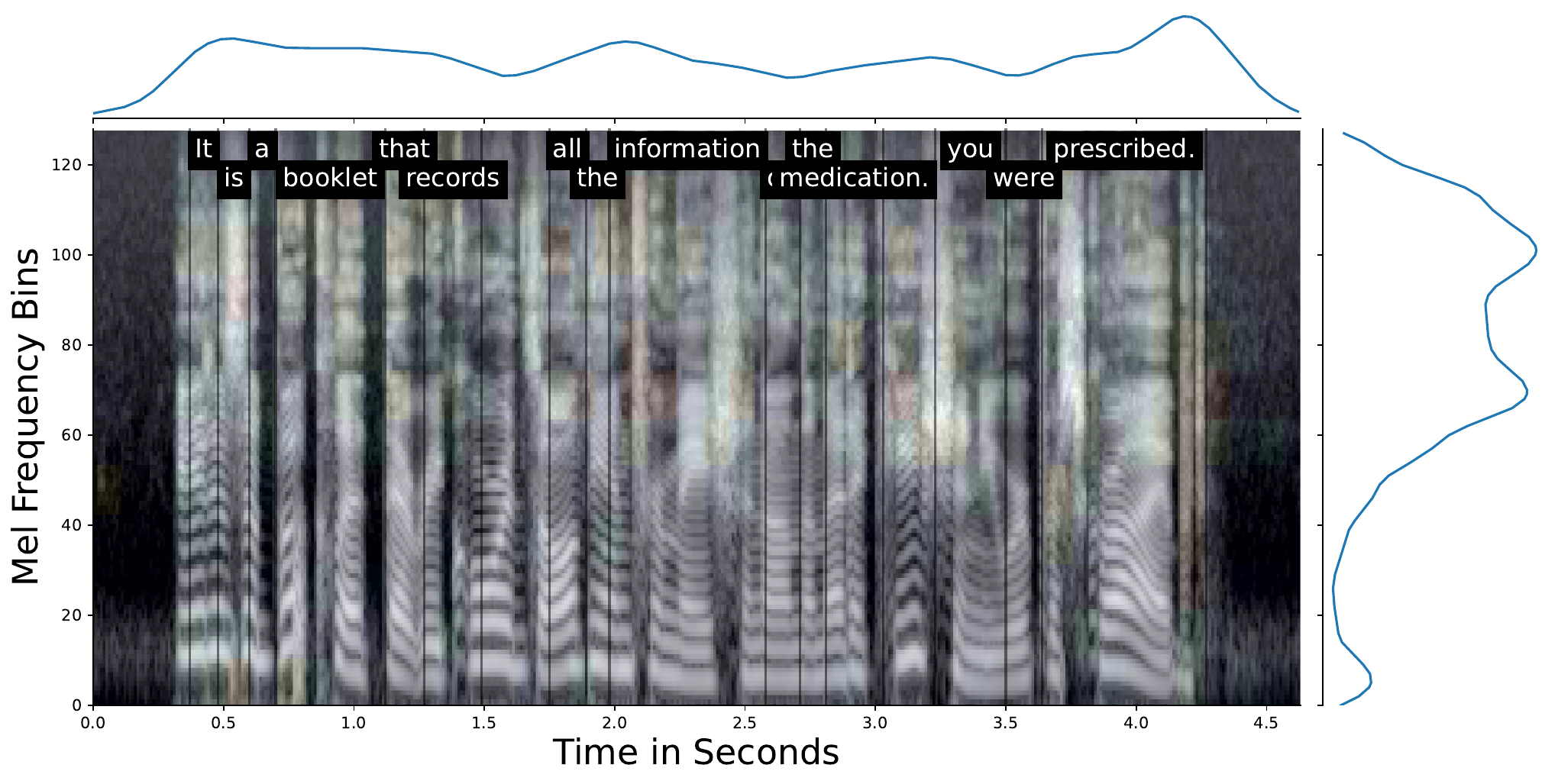}
    \caption{Example saliency output with KDE (in blue) and transcription temporally localising a discontinuity artifact on the last consonant at $4.2$s (sample from AudioMOS 2025~\cite{Huang2025}).}
    \label{fig:saliencyexample}
\end{figure}

Additionally the saliency maps are aggregated across the time and frequency domains using \gls{KDE} determining a smoothed distribution of the troublesome areas. These \gls{KDE} plots are shown against a waveform and transcription of the speech for explainability, as shown in Fig. \ref{fig:saliencyexample}. Using utterance level \gls{ASR} confidence, this value is plotted against the waveform and transcription, to show any words or phrases the system has struggled to transcribe. Measures on \gls{KDE} flatness across the time dimension, inspired by spectral~\cite{Madhu2009} and histogram~\cite{KumarTripathi2011} flatness, are also exported and added to the utterance metadata. This allows users to find the utterances where there are specific troublesome areas or general quality artifacts.

\subsection{System-Level Reports}

The output reports include two tsv files with a dump of the whole system \gls{SQ-AST} scores for each dimension requested, and the raw analysis values for the samples that lie under the threshold. Sorted tsv files are also provided for each dimension showing the lowest scoring samples first (along with other analysis details) allowing users to investigate utterance level reports by order of severity.

Reports that give a visual aid in showing the distributions of each \gls{SQ-AST} dimension are given as violin plot and bar chart indicating the amount of samples that fall below the threshold. The system-wide frequency aggregated \gls{KDE} is shown indicating the troublesome frequency bands for a given system. Additionally, correlation heatmaps between dimensions are shown for the whole dataset and the below-threshold dataset, to give an overview of inter-dimension trends. \gls{ASR} confidence intervals are shown as a distribution of utterance means and medians, indicating ease of transcription.

Histograms of system-wide troublesome phoneme pairs are shown as described in~\cite{Gupta2024}, capturing artifacts occurring at specific phoneme boundaries.

\section{Listening Test Design}

A listening test was designed in order to validate the efficacy of our pipeline in regards to the temporal extraction of troublesome features. Participants were asked to highlight a transcribed waveform for areas of the speech that they deem to be troublesome~\cite{seebauer2023re, Kuhlmann2025} for a given dimension. For each sample in the test, the results from all the participants were aggregated and normalised per sample. The discretised data is used to conduct \gls{PDSM} analysis, while applying \gls{KDE}, using the same bandwidth as the model, is used to detail agreement in the \gls{KDE} output of the system.

The synthetic speech used in the test comprised of 30 handpicked samples from the VoiceMOS 2022 datasets~\cite{Huang2022, Cooper2021}. The samples used have localized temporal artifacts, associated with low \gls{KDE} flatness scores, along with variance in the synthesis methodology and scoring for each dimension. The test was designed using Psychopy~\cite{peirce2019psychopy2} and run online with a headphone screening method~\cite{Woods2017} to ensure adequate responses.

\section{Results}
\label{sec:results}

The listening test was completed by $40$ participants, comprised of $20$ female, $19$ male and $1$ other participant. The mean age was $32.8$ years ($\mathrm{SD}~=~8.1$, range $24$-$55$). No diagnosed hearing impairments were reported by participants.

\subsection{SQ-AST Score Validation}

\begin{figure}[t!]
    \centering
    \includegraphics[width=1.0\linewidth]{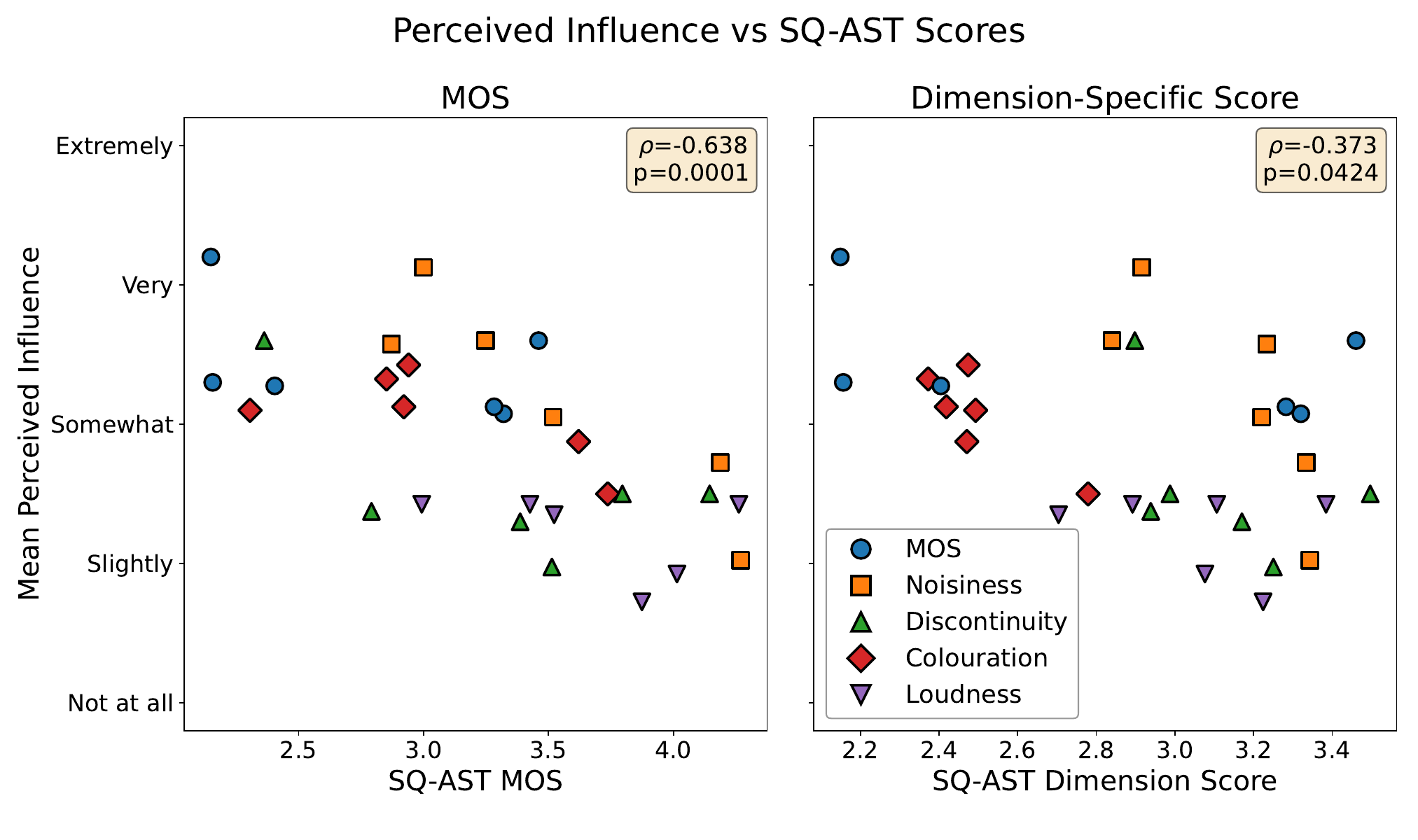}
    \caption{Perceived influence versus SQ-AST scores. Each marker corresponds to one of the listening test items.}
    \label{fig:influence_scores}
\end{figure}

Before assessing temporal localisation, the relationship between \gls{SQ-AST} scores and perceived artifact severity was examined. As expected, a significant negative correlation was found between mean perceived influence ratings and \gls{SQ-AST} \gls{MOS} values (Spearman's $\rho~=~-0.638$, $p~=~0.0001$). Lower \gls{SQ-AST} \gls{MOS} values thus corresponded to higher listener-reported annoyance from artifacts. \gls{SQ-AST} dimension-specific scores were also correlated with perceived influence, though more weakly ($\rho~=~-0.373$, $p~=~0.042$). The relationship is illustrated in Fig.~\ref{fig:influence_scores}.

\subsection{Temporal Artifact Localisation}

Temporal localisation was evaluated by comparing the \glspl{KDE} derived from each saliency method with the \gls{KDE} derived from aggregated listener highlight annotations. Spearman's $\rho$ was used to measure rank agreement between model and listener attention over time. The methods considered were Raw Attention, Attention Rollout, Attention Flow, \gls{GradCAM}$+$ (positive) and \gls{GradCAM}$-$ (negative). Attention Rollout and Attention Flow propagate attention through the \gls{AST} layers of the \gls{SQ-AST} model. Table~\ref{tab:localisation} reports the median sample-wise correlation for each method and quality dimension.

\begin{table}[t]
\centering
\small
\setlength{\tabcolsep}{4pt}
\caption{Median Spearman's $\rho$ between model and listener KDEs. The highest value per dimension is shown in bold.}
\label{tab:localisation}
\begin{tabular}{p{1.3cm}S[table-format=-1.3,detect-all=true]S[table-format=-1.3,detect-all=true]S[table-format=-1.3,detect-all=true]S[table-format=-1.3,detect-all=true]S[table-format=-1.3,detect-all=true]}

\toprule
\multicolumn{1}{c}{\rotatebox{45}{Dimension}} & 
\multicolumn{1}{c}{\rotatebox{45}{Raw}} & 
\multicolumn{1}{c}{\rotatebox{45}{Rollout}} & 
\multicolumn{1}{c}{\rotatebox{45}{Flow}} & 
\multicolumn{1}{c}{\rotatebox{45}{GradCAM$+$}} & 
\multicolumn{1}{c}{\rotatebox{45}{GradCAM$-$}} \\
\midrule

Overall & -0.020 & \bfseries 0.444 & 0.378 & 0.403 & 0.034 \\
MOS & 0.111 & 0.542 & 0.473 & \bfseries 0.649 & -0.521 \\
Nois. & -0.226 & 0.070 & \bfseries 0.389 & 0.380 & 0.155 \\
Discont. & 0.076 & \bfseries 0.358 & 0.234 & 0.257 & 0.017 \\
Colour. & -0.019 & 0.620 & \bfseries 0.732 & 0.088 & 0.089 \\
Loud. & -0.230 & 0.398 & 0.264 & \bfseries 0.625 & 0.041 \\
\bottomrule
\end{tabular}
\end{table}

\begin{table}[t]
\centering
\small
\setlength{\tabcolsep}{4pt}
\caption{Best method combination chosen by the maximum median Spearman's $\rho$ between model and listener \gls{PDSM}.}
\label{tab:pdsmbest}
\begin{tabular}{p{1.3cm}S[table-format=-1.3,detect-all=true]S[table-format=-1.3,detect-all=true]S[table-format=-1.3,detect-all=true]}
\toprule
\multicolumn{1}{c}{{Dimension}} & 
\multicolumn{1}{l}{{Best Method Combination}} & 
\multicolumn{1}{c}{{Spearman's $\rho$}} \\
\midrule
Overall & \multicolumn{1}{l}{GradCAM$+$ ($\ell_1$ norm, sum)} & 0.634 \\
MOS & \multicolumn{1}{l}{Rollout ($\ell_2$ norm)} & 0.469 \\
Nois. & \multicolumn{1}{l}{Rollout (median)} & 0.806 \\
Discont. & \multicolumn{1}{l}{Flow ($\ell_2$ norm)} & 0.560 \\
Colour. & \multicolumn{1}{l}{Flow ($\ell_2$ norm)} & 0.474 \\
Loud. & \multicolumn{1}{l}{Rollout (median)} & 0.806 \\
\bottomrule
\end{tabular}
\end{table}

The highest overall median correlation was observed for Attention Rollout ($\rho~=~0.444$), followed by \gls{GradCAM}$+$ ($\rho~=~0.403$) and Attention Flow ($\rho~=~0.378$). Wilcoxon signed-rank tests on the sample-wise correlations indicated that Attention Rollout and Attention Flow each displayed higher median correlations than Raw attention and \gls{GradCAM}$-$ ($p < 0.001$). \gls{GradCAM}$+$ showed a similar pattern ($p \leq 0.001$). \gls{GradCAM}$+$ yielded the highest values for \gls{MOS} ($\rho~=~0.649$) and loudness ($\rho~=~0.625$), Attention Flow for colouration ($\rho~=~0.732$) and noisiness ($\rho~=~0.389$), and Attention Rollout for discontinuity ($\rho~=~0.358$). This indicates that the different attention mechanisms are sensitive to different artifact classes.

At the per-sample level, positive correlations were observed in $83.3\%$ of samples for Attention Rollout, $86.7\%$ for Attention Flow and $83.3\%$ for \gls{GradCAM}$+$. With a threshold of $\rho > 0.3$, the corresponding values were $66.7\%$, $60\%$ and $63.3\%$. Peak alignment was also examined. The time bin with the highest model saliency fell within the top $50\%$ of listener-highlighted bins on $70\%$ of samples for Attention Rollout, $73.3\%$ for Attention Flow and $60\%$ for \gls{GradCAM}$+$.

Table \ref{tab:pdsmbest} presents the best saliency and pooling method combinations across each dimension for \gls{PDSM} median Spearman's $\rho$ between the model and listener outputs. The \gls{PDSM} pooling method was applied to the full model and listener discretised values, giving ranked phonemes. A threshold of $50\%$ was used, resulting in rank analysis between the most troublesome phonemes while avoiding low listener consensus areas. In the dimensions of \gls{MOS}, discontinuity and colouration, no other method combinations achieved a median \(\rho\) in the margin of \(0.05\) from the best combination, whereas in the other dimensions, there were multiple methods that achieve this margin. Additionally, all saliency extraction methods achieved overall median $\rho~>~0.5$ using the $\ell_1$ norm and sum pooling methods.

\subsection{Per-Participant Validation}

As aggregated \glspl{KDE} can smooth over disagreement between individual participants, a further analysis was performed on a per-participant basis. Each participant's binary highlight vector was compared against corresponding model-derived \glspl{KDE} using AUC-ROC. Of the $1200$ participant-sample pairs, $1061$ contained both highlighted and non-highlighted bins. Attention Rollout, Attention Flow and \gls{GradCAM}$+$ yielded median overall AUCs of $0.666$, $0.626$ and $0.662$, while Raw attention and \gls{GradCAM}$-$ were close to chance. Median AUCs for each dimension for the three best methods are shown in Table~\ref{tab:auc}.

\begin{table}[t]
\centering
\small
\setlength{\tabcolsep}{3pt}
\caption{Participant-level median AUC-ROC for discriminating highlighted and non-highlighted time bins. The highest value per dimension is shown in bold.}
\label{tab:auc}
\begin{tabular}{p{1.3cm}S[table-format=1.3,detect-all=true]S[table-format=1.3,detect-all=true]S[table-format=1.3,detect-all=true]S[table-format=1.3,detect-all=true]S[table-format=1.3,detect-all=true]}
\toprule
\multicolumn{1}{c}{\rotatebox{45}{Dimension}} & 
\multicolumn{1}{c}{\rotatebox{45}{Raw}} & 
\multicolumn{1}{c}{\rotatebox{45}{Rollout}} & 
\multicolumn{1}{c}{\rotatebox{45}{Flow}} & 
\multicolumn{1}{c}{\rotatebox{45}{GradCAM$+$}} & 
\multicolumn{1}{c}{\rotatebox{45}{GradCAM$-$}} \\
\midrule
Overall & 0.496 & \bfseries 0.666 & 0.626 & 0.662 & 0.468 \\
MOS & 0.552 & \bfseries 0.691 & 0.640 & 0.686 & 0.375 \\
Nois. & 0.465 & 0.570 & 0.659 & \bfseries 0.665 & 0.610 \\
Discont. & 0.522 & 0.542 & 0.467 & \bfseries 0.628 & 0.449 \\
Colour. & 0.503 & \bfseries 0.741 & 0.737 & 0.501 & 0.576 \\
Loud. & 0.431 & 0.690 & 0.505 & \bfseries 0.861 & 0.340 \\
\bottomrule
\end{tabular}
\end{table}

\section{Conclusion}

\gls{XSQ-AST} provides a post-hoc framework for localising artifacts in synthetic speech without retraining. A listening-test conducted on 40 participants confirmed that \gls{SQ-AST} scores reflect perceived artifact severity, and that model-derived temporal \glspl{KDE} align with listener highlight annotations at both the aggregate and participant level. Notably, different saliency mechanisms appear to be best suited to different perceptual dimensions. \gls{PDSM} analysis further demonstrated that phoneme-level localisation is achievable with appropriate pooling strategies.

Future work will scale the evaluation to larger datasets, incorporate prosodic analysis via appropriate models (e.g., WavLM~\cite{Chen2022}), and refine phoneme-level localisation accuracy. The results demonstrate that post-hoc saliency from pre-trained speech quality models can yield interpretable, temporally localised diagnostics for synthetic speech artifacts.

%\section{Acknowledgments}
%This work was supported by CoSTAR (Convergent Screen Technologies and Performance in Realtime) Live Lab, funded by the Arts and Humanities Research Council, grant reference AH/Y001079/1.

\clearpage
\bibliographystyle{IEEEbib_et_al}
% \small

\bibliography{strings,refs}

% Footnote: funding ack. + EA disclaimer
{\renewcommand{\thefootnote}{}\footnotetext{
The authors/contributors from Electronic Arts (EA) collaborated on this academic research project through supervision of the work. The views, conclusions, methods, and results presented do not necessarily represent the official position, technical claims, product plans, or future direction of EA.}

%This work was supported by EPSRC (Engineering and Physical Sciences Research Council) Impact Accelerator award EP/X525856/1 and by CoSTAR (Convergent Screen Technologies and Performance in Realtime) Live Lab, funded by the Arts and Humanities Research Council, grant reference AH/Y001079/1.

\end{document}